\documentclass[11pt]{article}

\usepackage{fullpage}
\usepackage{amsmath}
\usepackage{amssymb}
\usepackage{mathrsfs}
\usepackage{gensymb}
\usepackage{setspace}
\usepackage{bbm}
\usepackage{dsfont}
\usepackage{float}
\usepackage{subfigure}
\usepackage{adjustbox}
\usepackage{sidecap}
\usepackage{natbib}
\AtBeginDocument{}
\usepackage[hyphens]{url}

\usepackage{color,soul}

\setcitestyle{authoryear,open={(},close={)}}
\setcitestyle{citesep={,}}
\setcitestyle{aysep={}}

\usepackage{graphics}
\usepackage{authblk}
\usepackage[latin1]{inputenc} 
\usepackage{color}
\usepackage[american]{babel}
\usepackage{lineno}
\usepackage{setspace} 
\usepackage[colorlinks=true]{hyperref} 
\hypersetup{
    bookmarks=true,         
    unicode=false,             
    pdftoolbar=true,           
    pdfmenubar=true,        
    pdffitwindow=false,      
    pdfstartview={FitH},     
    pdftitle={My title},         
    pdfauthor={Author},     
    pdfsubject={Subject},   
    pdfcreator={Creator},   
    pdfproducer={Producer},    
    pdfkeywords={keyword1} {key2} {key3},  
    pdfnewwindow=true,     
    colorlinks=true,              
    linkcolor=blue,               
    citecolor=red,                
    filecolor=magenta,         
    urlcolor=cyan                 
}

\newcommand{\be}{\begin{equation}}
\newcommand{\ee}{\end{equation}}
\newcommand{\bes}{\begin{equation*}}
\newcommand{\ees}{\end{equation*}}
\newcommand{\bea}{\begin{eqnarray}}
\newcommand{\eea}{\end{eqnarray}}
\newcommand{\beas}{\begin{eqnarray*}}
\newcommand{\eeas}{\end{eqnarray*}}

\newcommand{\bmat}{\begin{bmatrix}}
\newcommand{\emat}{\end{bmatrix}}

\title{The effect of global warming on Western Mediterranean seagrasses: towards an agent-based modelling approach}

\author[1]{Eva Llabr{\'e}s\thanks{evallabres@ifisc.uib-csic.es}}
\author[2]{Aina Blanco-Magad{\'a}n}
\author[2]{Marta Sales}
\author[1]{Tom{\`a}s Sintes}

\affil[1]{Institute for Cross-Disciplinary Physics and Complex Systems, IFISC (CSIC-UIB), Universitat de les Illes Balears, E-07122 Palma de Mallorca, Spain}
\affil[2]{Observatori Sociambiental de Menorca, Institut Menorqu{\'i} d{'}Estudis, 07702 Ma{\'o}, Spain}

\begin{document}

\maketitle

\abstract{
The Mediterranean Sea exhibits rapid warming rates compared to the global average leading to worrying consequences for its inhabiting organisms. Seagrasses are key structural elements in coastal ecosystems, and studying how temperature affects these species is crucial to anticipate the implications of global warming. In this work, we use an empirically-based numerical model to study the combined dynamics of {\it Posidonia oceanica} and {\it Cymodocea nodosa} and their resilience to sea warming. The model is parametrised using seagrass growth rates measured at the Western Mediterranean Sea.
Under favorable growth conditions, our simulations predict the emergence of a coexistence region at the front between mono-specific meadows. This region can be characterised by its width and local shoot densities, which are found to depend on the coupling parameter between {\it Posidonia oceanica} and {\it Cymodocea nodosa} species. 
Such regions have been empirically observed in Ses Olles de Son Saura (Balearic Islands, Western Mediterranean Sea).  A comparison between the field measurements at the study site with the model predictions has been used to fit the value of the coupling parameter.
Field data also relates the width of the coexistence region to the average length of {\it Posidonia oceanica} leaves at the front. Remarkably, a linear relationship is found between the coupling parameter and the leaf length. 
%
 %
%
In the presence of sea warming, the model predicts an exponential decay in the population of {\it Posidonia oceanica}, which is highly sensitive to temperature. This behaviour is a direct consequence of the clonal nature of the plant and can be characterised by the model parameters. Considering a scenario of high greenhouse emissions, our model forecasts that {\it Posidonia oceanica} meadows will lose 70$\%$ of their population by the year $2050$. {\it Cymodocea nodosa}, with higher thermal resilience, acts as an opportunistic species conquering the space left by the degraded {\it Posidonia oceanica}. 
}

\newpage

\section{INTRODUCTION}\label{sec:intro}

Ocean warming is one of the major causes behind the deterioration of many marine ecosystems \citep{Bruno2010}.
Especially alarming is the case of the Mediterranean Sea, which warms up three times faster than the ocean average due to its enclosed nature \citep{Vargas2008}. Other anthropogenic pressures, such as coastal development, water pollution, or increased mooring activity, are transforming the Mediterranean Sea into one of the most threatened marine areas, with part of its inhabiting species at high risk of declining. \citep{Templado2014}. This is the case of the endemic seagrasses {\it Posidonia oceanica}, which has undergone a considerable regression in past decades \citep{boudou,MARBA2014183,BURGOS20171108}. {\it P. oceanica}, together with other seagrasses species, is one of the most valuable elements in marine coastal areas since they provide ecosystemic services to many fishes and invertebrates. They also create architectural structure as benthic producers, contribute to the water quality and sediment stabilisation, protect coasts from strong waves, and are responsible for significant carbon sequestration \citep{Costanza_1997,bg-2-1-2005,Orth_2006,S_nchez_Gonz_lez_2011}. The loss of these ecosystems would be particularly devastating for people in coastal communities, where seagrass beds support local businesses and provide food security from fish production \citep{Watson_1993}.
\\
\\
The endemic {\it P. oceanica} shares the Mediterranean coastlines with the native seagrass species {\it Cymodocea nodosa}. Despite their common importance in providing ecosystemic services, these two seagrasses occupy different positions in the hierarchy of foundational species due to their distinct morphological features and ecological strategies \citep{mazzella}. 
{\it P. oceanica} forms robust meadows with thick, long-living rhizomes that extend both vertically and horizontally. Its leaves commonly have $1m$ length and generate a significant amount of biomass. As a consequence, {\it P. oceanica} has a slow growth, and its colonisation strategy mainly relies on shoot cloning, with sexual events irregular and uncommon \citep{PergentMartini_1994}. This is in contrast with {\it C. nodosa}, which has much thinner and shorter leaves, and a simpler rhizomatic structure lacking vertical growth.  Also,  {\it C. nodosa}  possesses fast colonisation patterns, compared to the slow rhizome growth rate of {\it P. oceanica} \citep{10.2307/24857076}, and has regular flowering events every year. 
The morphological characteristics of the plants make {\it P. oceanica} a more efficient ecosystem engineer than {\it C. nodosa}, with enhanced benthic production, a larger associated animal community, and a higher carbon burying capacity \citep{Marba1996,DUARTE1999159}. However, {\it C. nodosa} has an increased response to disturbances and a superior capacity for adaption due to its colonisation strategies \citep{Hughes_2004}, while {\it P. oceanica} is extremely vulnerable to changes in environmental conditions. These two species also differ in their thermal tolerances, while  {\it P. oceanica} meadows will dramatically decline for seawater temperatures higher than 29$\celsius$, {\it C. nodosa} optimally grows up to at least 34$\celsius$ \citep{Savva_2018}. Therefore, {\it C. nodosa} is expected to have higher resilience to warming events, probably due to the tropical origin of its genus \citep{Mar_n_Guirao_2016}. 
\\
\\
Studying the combined dynamics of {\it P. oceanica} and {\it C. nodosa} in response to thermal stress is essential to predict habitat shifts due to the climate change. Mathematical models offer a theoretical framework to investigate the relevant mechanisms that lead to the growth of seagrass ecosystems. Moreover, they can be used to evaluate the resilience to stress factors and identify tipping points where a small change in the environmental conditions may cause an irreversible loss of the population. Therefore, these models become a useful tool to assess the health of ecosystems and generate more informed policies to conserve marine coastal areas. 
 In \cite{TS2005}, they develop the first agent-based mathematical model that reproduces the spatial dynamics of seagrass meadows. Iterating a set of empirically-based clonal growth rules on every shoot, the model generates the non-linearities observed in the growth of young patches \citep{Sintes_2006}. This approach has proven useful in a broad range of applications, such as assessing the potential of seagrasses as carbon sinkers \citep{CDuarteCO2_2013} or estimating the age of the meadows  \citep{Arnaud_Haond_2012}. In \cite{Ruiz-Reynes:2017aa}, they propose a macroscopic description of the agent-based model using a set of partial differential equations for the shoot density. By losing detailed information on the rhizome structure, this model performs successfully at larger spatial scales and serves to explain the formation of patterns in seagrass beds \citep{Ruiz_Reyn_s_2019, Ruiz_Reyn_s2_2020}.  
 While previous models concentrated on the development of single species, in \cite{llabres2022mathematical} they extended the agent-based model to describe the interactions between seagrass species and other macrophytes. Inter-species interactions were introduced via competitive and facilitative relations that included a coupling parameter that could be tuned to reproduce field observations.
This multi-species model offers an innovative tool to understand interaction mechanisms between different species, evaluate the competition for resources between native and invasive species, and predict the spatial redistribution of the seagrasses after suffering from stress conditions or extreme environmental events. Understanding the response to thermal stress becomes crucial to anticipate the impacts of climate change.
\\
\\
In this article, we study the combined dynamics of {\it P. oceanica} and {\it C. nodosa}, focusing on their response to global warming. Their different thermal limits \citep{Savva_2018} suggest that future elevated water temperatures will cause a species shift favouring {\it C. nodosa} and drastically affecting habitat conditions in the Mediterranean Sea. We test this hypothesis using the agent-based numerical model for multi-specific seagrass ecosystems proposed in \cite{llabres2022mathematical}. Our simulations consider seagrass growth parameters directly measured from fieldwork performed at different sites in the Western Mediterranean Sea, previously existing in the literature and summarized in Table S1 in the Supplementary Material. 
We propose a novel methodology to determine the coupling coefficients between {\it P. oceanica} and {\it C. nodosa}. These coefficients measure the influence of one species upon another and cannot be obtained directly from field observations. 
Our approach involves the characterization of the front properties between the two mono-specific meadows, namely the front width and the shoot densities, and comparing the model outcome to the field data obtained in Ses Olles de Son Saura (Balearic Islands, Western Mediterranean Sea). Once the model parameters are fixed, we can study the temperature dependence of the combined dynamics of {\it P. oceanica} and {\it C. nodosa}. We have used the averaged Mediterranean warming trends predicted by \cite{Darmaraki} for different greenhouse gas emissions scenarios.

\section{MATERIALS AND METHODS}

\subsection{Numerical model}\label{sec:num}

\begin{figure}[H]\centering
\includegraphics[width=0.95\textwidth]{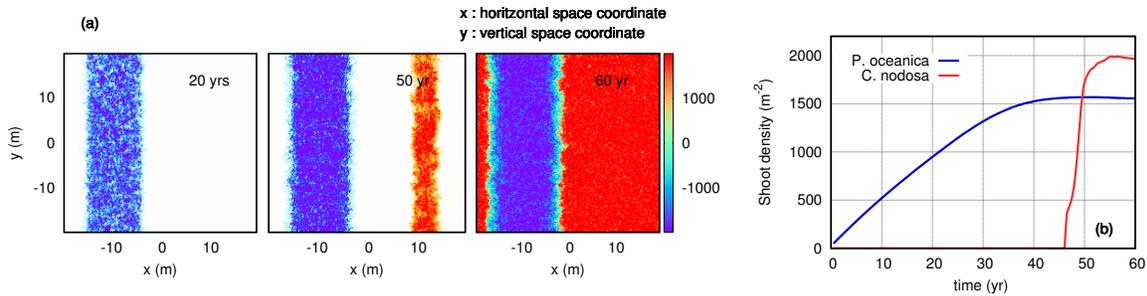}
    \caption{ \label{fig:CI} \small Set of the initial condition considered in our simulations. The shoots painted in light green represent  {\it Cymodocea nodosa} seagrass, and dark green shoots correspond to {\it Posidonia oceanica}. (a) Different snapshots of the system are taken at times $t=20,\,50,\,60\,yr$, whose color bar represents $\rho_{Po}-\rho_{Cn}$, the difference between shoot densities of  {\it P. oceanica} ($\rho_{Po}$) and {\it C. nodosa} ($\rho_{Cn}$) in units of [shoots$\cdot m^{-2}$]. Regions coloured in blue (red) are dominated by the presence of {\it P. oceanica} ({\it C. nodosa}). (b) Change in time of the average shoot densities of both species.
    }
 \end{figure}

The model follows a set of growth rules that, based on the clonal behavior of the plant, that are related to the following empirical parameters: the rhizome elongation rate $[v]$, which sets the horizontal spread of the clone; the branching rate $[\nu_0]$, that controls the capacity of the clone to form dense networks; the branching angle $[\phi]$, that determines the efficiency of the space occupation; the spacer length $[\delta]$, that measures the length of the rhizome between consecutive shoots; and the shoot mortality rate $[\mu]$ \citep{BELLTOMLIN,Marb__1998}. For more details on the implementation of the model, see Sec. SM1 in the Supplementary Material. The values of the clonal growth parameters are collected from direct field observations performed at various sites of the Western Mediterranean region summarized in Table S1, in the Supplementary Material. The input parameters are characterized by their average and standard deviation and implemented at each time step of the model following a Gaussian distribution.
\\
\\
The model incorporates a shoot density-dependent branching rate \citep{llabres2022mathematical}, i.e. $\nu(\rho)=\nu_0+\epsilon \hat\rho \left (1-\hat\rho \right )$, where  $\hat\rho = \rho/\rho_{max}$ is the normalised local density, and $\epsilon$ is a coefficient that controls the species carrying capacity and facilitation terms. 
The parabolic shape of $\nu(\rho)$ penalises over- and under-populated areas and favours regions around an optimal density  $\rho = \rho_{max}/2$. 
The coupling between the two different species ($1\,\&\,2$) is included in the definition of the local shoot density as $\hat \rho_i={\left(\rho_i + \gamma_{ij} \rho_j\right)/\rho_{max,i}}$, where $\gamma_{ij}$ is the coupling coefficient and $i\neq j = 1,2$. The higher the value of $\gamma_{12}$, the more seagrass species $1$ is affected by the presence of species $2$, and vice-versa. The coupling coefficients $\gamma_{ij}$ cannot be fixed by direct measurements and must be inferred indirectly by comparing the solutions of the model to field observations. In \citep{llabres2022mathematical} two possible regimes were found:
\begin{itemize}
\item[-]  \textbf{Mixed meadows:} Stable mixed meadows develop when the coupling parameters satisfy the inequalities: $\gamma_{12} < 1$ and $ \gamma_{21} < 1$. In this case, self-interactions are higher than inter-specific ones and the competition for the available space is minimized when the shoots of different species mix.
\item[-] \textbf{Mono-specific meadows:} The condition $ \gamma_{12} > 1$  ($ \gamma_{21} > 1$) favours the self-interaction between individuals of species 1 (2)  and causes species to separate into mono-specific domains. 
 \end{itemize} 
In the present work, we have analysed the interaction between {\it P. oceanica} (1) and {\it C. nodosa} (2), the two most relevant seagrasses species in the Mediterranean Sea. {\it P. oceanica} has a strong rhizomatic structure and much longer and thicker leaves than {\it C. nodosa} \citep{mazzella}. Therefore, we can assume that  {\it P. oceanica} grows almost unaffected by the presence {\it C. nodosa}. In all simulations, we set  $\gamma_{12}=0.1$ and have explored the response of the system to different values of $ \gamma_{21} > 1$. This combination of coupling parameters will result in domain-separated solutions of {\it P. oceanica} and {\it C. nodosa}, as expected. 
The optimal densities for each species are selected to be $\rho_{max,1}/2 = 900$ shoots$ \cdot m^{-2}$ and $\rho_{max,2}/2 = 1200$ shoots$ \cdot m^{-2}$. With these values, our simulations will reproduce healthy and realistic meadows with densities similar to those measured in the Western Balearic Sea \citep{Marb__1996,informeMarBalear}. We also set $\epsilon_i = \nu_{0,i}/2$, such that the effective branching rates $\nu_i(\rho)$ are consistent with field observations. 
\\
\\
The effect of the seawater temperature on seagrasses differs from one species to another \citep{Savva_2018}. In \citep{marba_warming}, experimental measurements performed in Cabrera National Park (Balearic Islands, Western Mediterranean Sea) indicate that natural populations of {\it P. oceanica} increment their mortality rate linearly as $\mu \approx 0.028  \, SST_{max}$, where $SST_{max}$ is the yearly maximum of the surface seawater temperature. 
The averaged value of $SST_{max}$ is also expected to increase almost linearly in time \citep{Darmaraki}, i.e.,  $SST_{max} \approx \lambda  t$, where $\lambda$ is the increase rate of the seawater temperature $[\celsius\cdot yr^{-1}]$. Therefore, the mortality rate of {\it P. oceanica} in the presence of warming can be written as: 
	\be\label{mortt}
	 \mu(t) = \alpha   \,t + \mu_0\,
	 \ee
where $ \alpha = 0.028 \, \lambda$, and the value of the parameter  $\lambda$ can be tuned to model different future warming scenarios. The intrinsic shoot mortality rate, $\mu_0$, will also account for the local anthropogenic pressures and is taken to be $\mu_0 =  0.07 yr^{-1}$. On the contrary, {\it C. nodosa} is not negatively affected by seawater temperatures below $34\celsius$ as shown by \cite{Savva_2018} with experiments performed with specimens from different sites in Mallorca (Balearic Islands, Western Mediterranean Sea), and we assume its mortality rate to be constant through this work. We will start our simulations setting  $SST_{max,0} = 29\celsius$, the maximum surface seawater temperature averaged over different stations in the Balearic Sea region in the year 2020 \citep{informeMarBalear2}.
We have performed simulations using the predictions found in \citep{Darmaraki}, with the following sea-warming rates:  $\lambda= 1.0,\,2.0,\,4.5\,\celsius/100\,yr$. These trends correspond to Mediterranean averages, and are found using a multi-model approach in three different Representative Concentration Pathways of greenhouse gas  (RCP2.6, RCP4.5, RCP8.5) taken from the {\it IPCC Fifth Assessment Report} \citep{IPCC2014}. Scenario RCP2.6 corresponds to stringent mitigation measures that aim to keep global warming below 2$\celsius$ above the pre-industrial level, RCP4.5 represents a positively-moderate case, and RCP8.5 is a scenario with very high greenhouse gas emissions. 
\\
\\
In our simulations, we consider a representative region of space with an area of $L \times L$,  with $L=  20 \, m $, and periodic boundary conditions.
 A square grid, representing the experimental transects, is superimposed on top of the continuum space. The cells on the grid have a size of  $20 \times 20 \, cm^2$, resulting in a total of $10000$ cells. The position and the number of shoots populating each cell are known and are used to estimate the average shoot density and other relevant magnitudes. To account for the variability in temperature predictions and growth parameters, we have averaged all the numerical results presented in this work over 15 independent realisations.
\\
\\
Since the growth rate of {\it C. nodosa} is much higher than that of  {\it P. oceanica}, 
in order to generate stable meadows of comparable sizes for both species as an initial condition before the increase of the seawater temperature is evaluated,  we proceed as follows (see Fig. \ref{fig:CI}): 
$i)$ The simulations start with a $1m$ wide vertical stripe of homogeneously distributed {\it P. oceanica} seeds located at $x=-10\,m$. 
A patch of {\it P. oceanica} develops until it reaches its stable density at $t=46\, yr$ (Fig  \ref{fig:CI} (right)). $ii)$ At this time we add a similar stripe of {\it C. nodosa} located at $x=10\,m$. The striped meadows keep growing wider with time until they encounter each other at $t=60\, yr$. At this point, the percentage of the total coverage is $40\%$ for  {\it P. oceanica} and $60\%$ for  {\it C. nodosa}. This configuration is used as the initial condition to evaluate the effect of the sea surface warming.

\qquad

\subsection{Experimental design}\label{sec:exp}

\par

\begin{figure}[H]\centering
\includegraphics[width=0.6\textwidth]{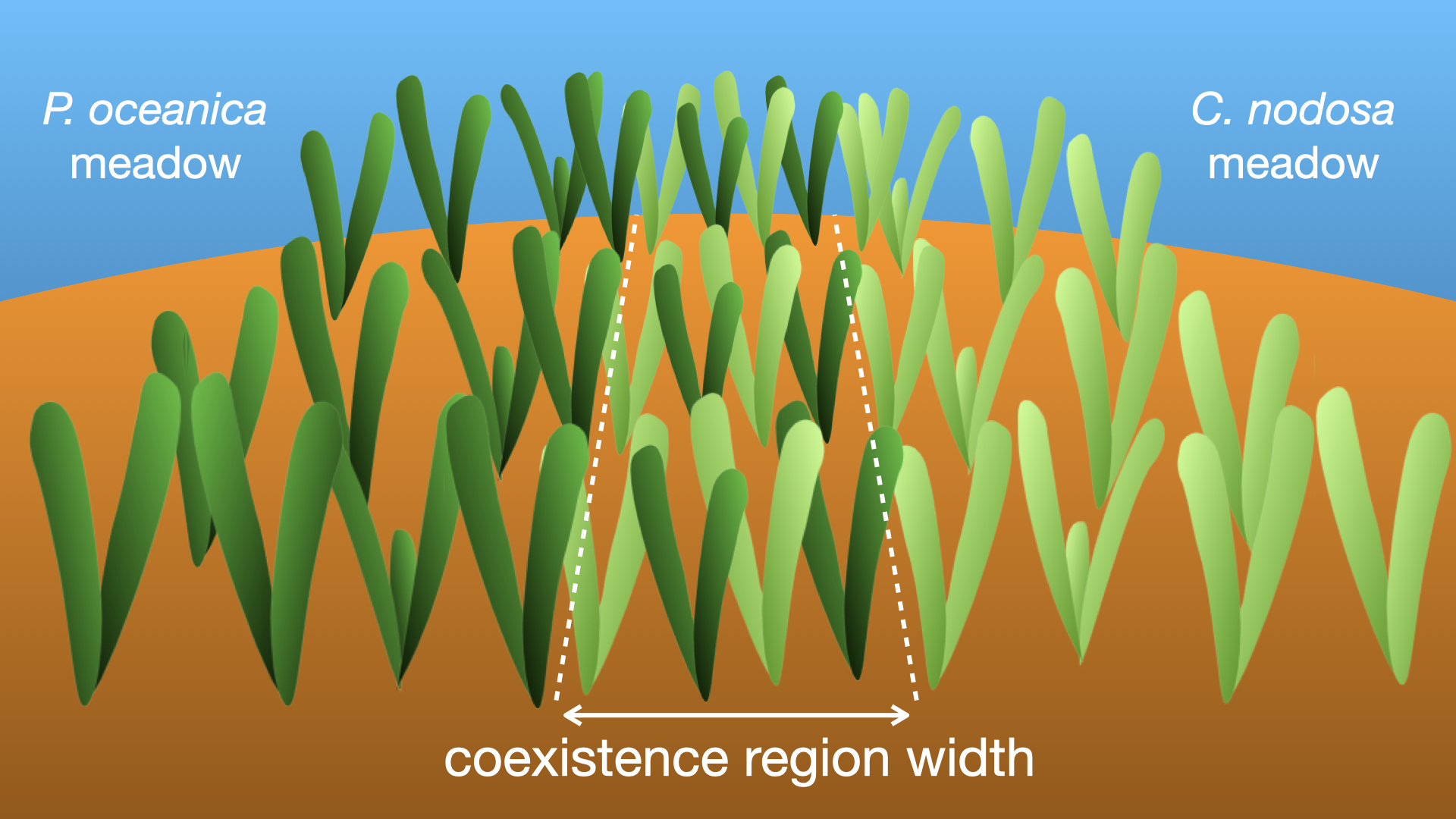}
    \caption{ \label{fig:draw} \small Schematic drawing of the seagrass patterns observed in ses Olles de Son Saura (Balearic Islands, Western Mediterranean Sea). The shoots painted in light green represent  {\it Cymodocea nodosa} seagrass, and dark green shoots correspond to {\it Posidonia oceanica}. The region marked with the white dotted lines corresponds to the stripe-like coexistence regions observed at the front between mono-specific meadows of  {\it C. nodosa} and  {\it P. oceanica}. The white arrow exemplifies the width field measurements performed at the study site, selecting the distance from the right outer-most shoot of {\it P. oceanica} to the left outer-most shoot of {\it C. nodosa}.}
 \end{figure}

In this work, we propose a novel method to quantify the influence of {\it P. oceanica} over {\it C. nodosa}, tuning the coupling parameter $ \gamma_{21}$. Our numerical model predicts the emergence of a coexistence region at the front between mono-specific meadows that is characterized by its width and local shoot densities. The observation and measurement of such regions in the study site allow us to fit the coupling coefficient $ \gamma_{21}$.
 The details of the experimental design are specified in the following.
 \\
\\
Field measurements were conducted in Ses Olles de Son Saura, located on the north coast of Menorca (Balearic Islands, Western Mediterranean Sea) in October 2021. The site is a sheltered cove with an estimated surface area of $0.3\, km^2$. The fieldwork was performed in areas with a maximum depth of $2\,m$ that were covered by the seagrasses {\it P. oceanica} and {\it C. nodosa}. The two species, organised in mono-specific meadows, occupy different domains of the seafloor. However, we could often identify a narrow and well-defined {\it coexistence region} at the frontier between meadows which had the appearance of a stripe with a seemingly regular width (Fig. \ref{fig:draw}). A minimum of five different coexistence regions were analysed in the study site.  We measured the front width, the {\it C. nodosa} shoot density and the {\it P. oceanica} leaf length in each location. The shoot densities were measured by counting the number of shoots in a $20\times 20 \, cm^2$ quadrant. Points with less than seven shoots of {\it P. oceanica} or {\it C. nodosa} were not considered. We also randomly selected seven {\it P. oceanica} leaves per quadrant, measured their length from the ground to the tip, and performed their average. The width at each of the selected regions of coexistence was measured from the outer-most shoot of {\it P. oceanica} to the outer-most shoot of {\it C. nodosa} at each side of the stripe. 
\\
\\
We also measured the averaged {\it C. nodosa} shoot density in the mono-specific meadow at the location. The observed value of $1416\pm 652$ shoots$ \cdot m^{-2}$  is comparable to data collected in June 2021 of $1450\pm 447$ shoots$ \cdot m^{-2}$, during the seasonal peak in seagrass biomass and shoot density \citep{2014ECSS..142...23M}. This result indicates that our measurements would not be significantly different if they had been collected during the season of maximal growth of {\it C. nodosa} that takes place between June and August.

\section{RESULTS}
\label{sec:res}

In this section, we aim to examine the proposed numerical model and validate its results by direct comparison with field observations. First, will select those parameter values that allow the development of the seagrass species under favourable conditions. We will analyse the meadow properties and, more interestingly, the coexistence region at the fronts between mono-species domains. In the second part, a time-dependent mortality rate, following eq. \eqref{mortt}, will be incorporated in the model to predict the changes in the ecosystem for different warming rates of 
the seawater temperature.

\subsection{{\it P. oceanica} and {\it C. nodosa} growth under favorable conditions}
\label{sec:good}

\subsubsection{Numerical simulations of the coexistence region}\label{sec:front}

\begin{figure}[H]\centering
\includegraphics[width=1\textwidth]{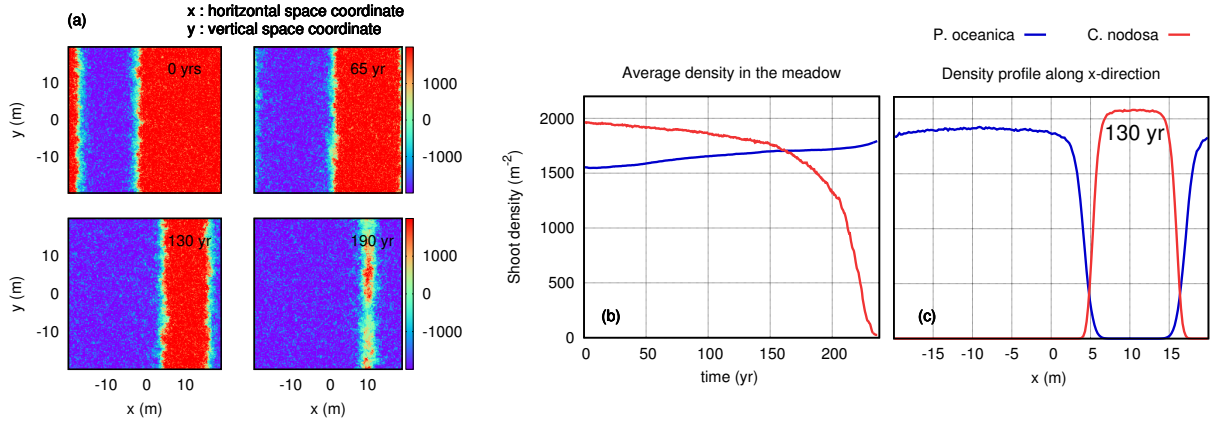}
    \caption{ \label{fig:general} \small (a) Domain separation of {\it Posidonia oceanica} and {\it Cymodocea nodosa} with coupling coefficients $ \gamma_{12}=0.1$ and $\gamma_{21}= 6$. Different snapshots are taken at times $t=0,\,65,\,130,\,190\,yr$, whose color bar represents $\rho_{Po}-\rho_{Cn}$, the difference between shoot densities of  {\it P. oceanica} ($\rho_{Po}$) and {\it C. nodosa} ($\rho_{Cn}$) in units of [shoots$\cdot m^{-2}$]. Regions coloured in blue (red) are dominated by the presence of {\it P. oceanica} ({\it C. nodosa}), and the thin green stripes represent the region of coexistence between both species. (b) Change in time of the average shoot density, and (c) the shoot density profile along the x-horizontal direction at $t= 130\,yr$.
    }
 \end{figure}

\begin{figure}[H]\centering
   {\includegraphics[width=0.9\textwidth]{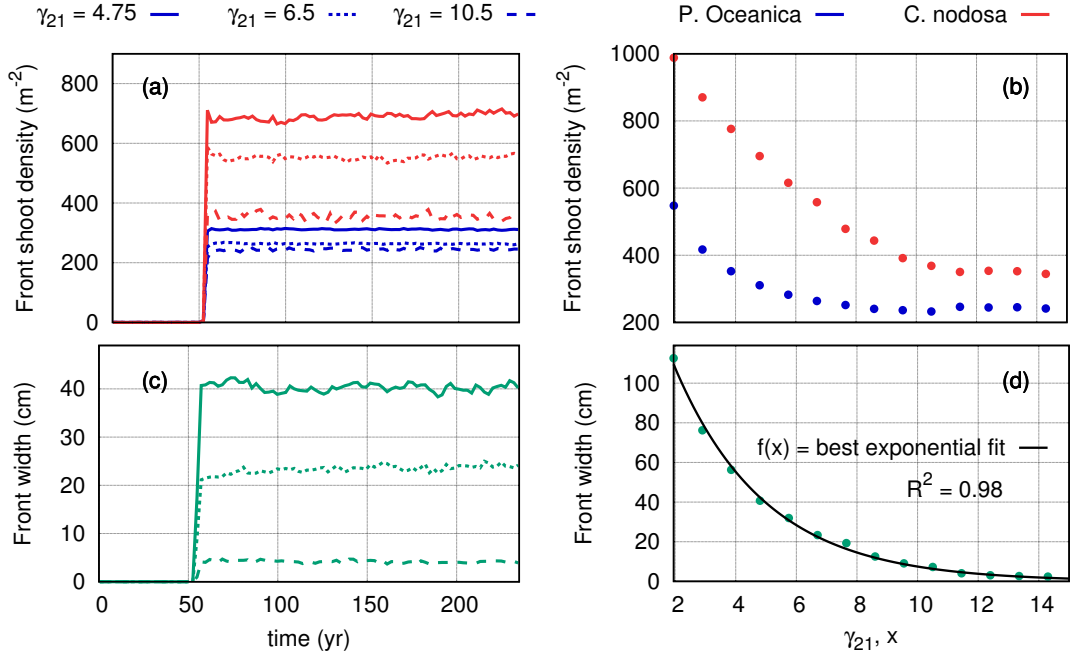}}
\caption{\label{fig:width} \small Our simulations show that {\it Posidonia oceanica} (red) and {\it Cymodocea nodosa} (blue) separate into clear spatial domains divided by a narrow stripe-like coexistence region at the front. The characteristics of this region depend on the influence that one species has over each other, which is represented by the coupling coefficients $ \gamma_{21}$.  In this figure, we analyse: (a) the change in the shoot density at the coexistence region (or front) for selected values of the coupling coefficient $ \gamma_{21}$ , (b) the saturation shoot densities, also at the front, vs. $ \gamma_{21}$,  (c) the change in the average front width for selected values of $ \gamma_{21}$, and (d) the width of the front vs. the coupling coefficient $ \gamma_{21}$. The solid backline represents the best fit for an exponential decay function, which corresponds to $f(x) = (209\pm 5)\exp{(-0.33\pm0.01)x}$. The width is calculated using the formula ${\cal W} = r_c \, N_{coex}/2\,  N_y$, where $N_{coex}$ is the number of $20\times 20 \, cm^2$ grid cells where coexistence between species takes place, $N_y$ are the number of cells in a column along the $y$-axis and $r_c=20 cm$ is the cell edge size. As it is done during field measurements, cells with less than seven shoots of any species were not considered part of the coexistence region.}
\end{figure}

We simulate the combined dynamics of {\it P. oceanica} and {\it C. nodosa} growing under favorable environmental conditions and neglecting temperature effects. 
Our model shows that species separate into clear spatial domains divided by a narrow stripe-like coexistence region. The results of our simulations are summarised in Fig. \ref{fig:general}. We have fixed $\gamma_{12}=0.1$ and $\gamma_{21} = 6$. This choice of coupling parameters is motivated by the different anatomy of both species, which indicates that the influence of {\it C. nodosa} over {\it P. oceanica} is almost negligible \citep{mazzella}.
The condition $ \gamma_{21} \gg 1 \gg \gamma_{12}$ ensures that the dominant species is {\it P. oceanica}, that keeps on winning space over {\it C. nodosa} (Fig. \ref{fig:general} (left)). 
In Fig. \ref{fig:general} (right) (a), we plot the average shoot density of both species as a function of time. We observe that {\it C. nodosa} completely disappears after $t=300\, yr$. In Fig. \ref{fig:general}(right) (b), we plot the density profile along the spatial x-direction for both species at $t=190\, yr$. We observe that the average shoot densities are fairly constant at the interior of the mono-specific meadows and decrease abruptly when approaching the front. The overlap of both density curves indicates the presence of a region of coexistence at the boundaries. In the snapshots of Fig. \ref{fig:general} (left), this region is identified by the narrow green stripes located at the borders between the domains.
\\
\\
In Fig. \ref{fig:width}, we study the dynamics of the coexistence region at the front between  {\it P. oceanica} and {\it C. nodosa} for different values of the coupling coefficient $ \gamma_{21}>1$. Our main observation is that the width and shoot density at the coexistence region depends significantly on the coupling coefficient $\gamma_{21}$.
%
%
From Fig. \ref{fig:width}(a, c), we observe that the shoot densities and the width of the coexistence region remain constant over time. However, both quantities decrease with increasing values of $\gamma_{21}$. This effect is particularly relevant in the case of {\it C. nodosa} (Fig. \ref{fig:width}(a)). This result is expected since for increasing values of $\gamma_{21}$, shoots of {\it C. nodosa} are more easily expelled from the areas occupied by {\it P. oceanica}, and the coexistence region shrinks.
In Figs. \ref{fig:width}(b, d) we plotted the average shoot densities and the front width as a function of $\gamma_{21}$. 
A monotonic decrease is observed for both quantities: the higher the values of $ \gamma_{21}$, the more the density of {\it C. nodosa} is negatively affected by the presence of {\it P. oceanica}, and the narrower and less populated is the front. However, since shoot density values decrease at the front, {\it C. nodosa} can still invade a small region of the {\it P. oceanica} meadow. Thus, the coupling coefficient $\gamma_{21}$ dictates the value of the {\it P. oceanica} shoot density below which the coexistence with {\it C. nodosa} emerges.  
Our simulations also suggest an exponential decay of the front width with $ \gamma_{21}$. The best fit to the data is shown in Fig. \ref{fig:width}(d).

\subsubsection{Model predicitions: A comparison with field observations}\label{sec:comp}

\begin{figure}[H]\centering
   {\includegraphics[width=0.95\textwidth]{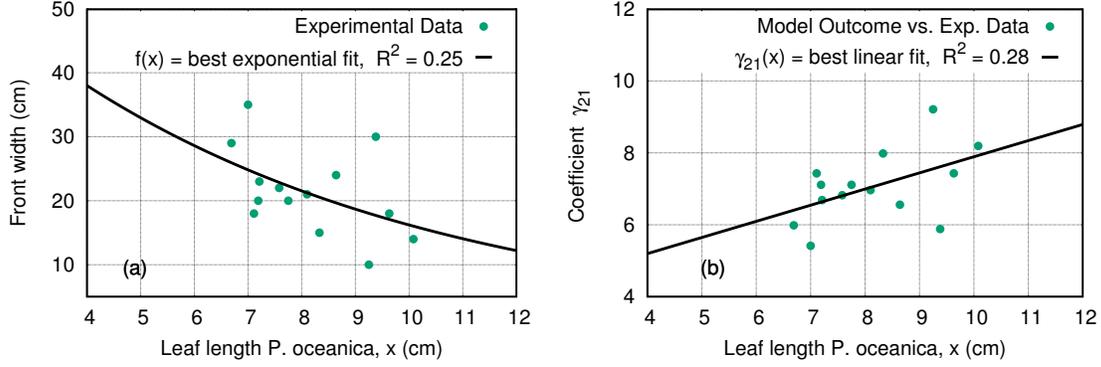}
   \caption{\label{fig:leafs} \small Experimental data collected at the coexistence fronts in Ses Olles de Son Saura (Balearic Islands, Western Mediterranean Sea). (a) Width of the coexistence region vs. the length of {\it Posidonia oceanica} leaves.  The solid black line represents the best fit to the experimental data assuming an exponential decay, which corresponds to $f(x) = (67\pm 40)\exp{(-0.14\pm0.08)x}$. (b) Relationship between the coupling coefficient of the model $ \gamma_{21}$ and the average leaf length of   {\it P. oceanica} shoots at the front measured at the study site. The best fit corresponds to the linear function: $\gamma_{12} = (0.45\pm0.2)x\,+\,(3.4\pm1.8)$. }
        }
     \end{figure}

One of our main objetives is to test the 
multi-specific numerical model with field observations performed in Ses Olles de Son Saura (Balearic Islands, Western Mediterranean Sea). 
At the study site, we have observed stripe-patterned coexistence regions between mono-specific meadows of {\it P. oceanica} and {\it C. nodosa} like those simulated in Sec. \ref{sec:front}. In Fig. \ref{fig:width}, we have seen the relationship between the coupling parameter $ \gamma_{21}$ and the width and densities at the simulated stripes. The experimental analysis of these stripes provides an estimated value for  $ \gamma_{21}$ in geographical areas with similar conditions to our sampling area. 
In the field, we measured the coexistence regions to have an average width of $21\pm 2\,cm$, and a {\it C. nodosa} shoot density of $613\pm70\,m^{-2}$. By direct comparison with Fig. \ref{fig:width}(b,d), these measurements suggest a value for the coupling parameter of $\gamma_{21}\approx 6.5$. 
 \\
 \\
We have also measured an average length of {\it P. oceanica} leaves at the coexistence region of $8.15 \pm 0.3\,cm$. This value is considerably lower than the expected one for {\it P. oceanica} in stable and healthy meadows  ($\sim 1m$) \citep{mazzella}. This observation may indicate that particularly short {\it P. oceanica} leaves are required for the emergence of coexistence fronts. 
In addition, we found a negative experimental relationship between the width of the front and the average leaf length of {\it P. oceanica}, which is best fitted by an exponential decay function (see Fig. \ref{fig:leafs} (a)). This result, together with the exponential fit between the front width and the coupling $ \gamma_{21}$ (Fig. \ref{fig:width}(d)), suggests an approximately linear relationship between the coupling coefficient $\gamma_{21}$ and the leaf length of {\it P. oceanica} (see Fig. \ref{fig:leafs} (b)). As a consequence, and this is a remarkable result, the size of the leaf length of {\it P. oceanica} can be linked to the coupling coefficient that mediates the interaction between  {\it P. oceanica} and {\it C. nodosa} at the coexistence region.

\quad

\subsection{Ecosystem dynamics under different seawater warming scenarios}\label{sec:temp}

\begin{figure}[H]\centering 
\includegraphics[width=0.99\textwidth]{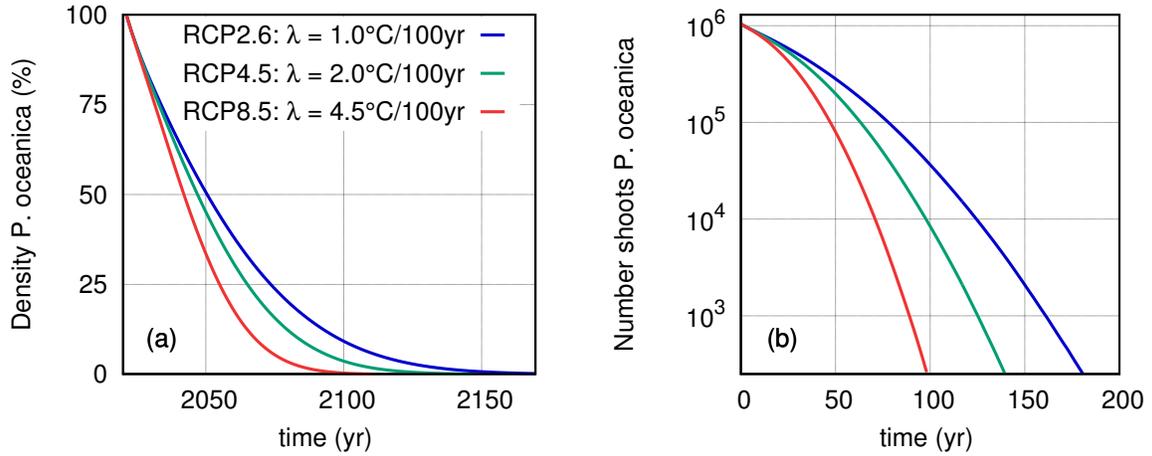}
\caption{\small\label{fig:exp} (a) Population decay (in $\%$) of {\it Posidonia oceanica} meadows subjected to different global warming scenarios corresponding to the Representative Concentration Pathways RCP2.6, RCP4.5, RCP8.5 from the {\it IPCC Fifth Assessment Report} \citep{IPCC2014}. The mortality rate of the seagrass increases linearly with time: $\mu(t) = \alpha  \,t\, + \mu_0$. The initial mortality rate is fixed to $\mu_0 = 0.07\,yr^{-1}$, the branching rate is set to  $\nu_0 = 0.05\,yr^{-1}$, and $\alpha = 0.028\lambda$, where $\lambda$ is the sea temperature increase rate. (b) Change in the number of shoots on a semi-logarithmic scale. The results are best fitted assuming a quadratic dependence on time: $f(t)=a{t^2} + b t + c$. The values of the parameters $a,b,c$ are summarised in Table \ref{table:exp}. The uncertainties of the different predictions are displayed in Fig. S1 in the Supplementary Material. }
 \end{figure}

\begin{figure}[H]\centering
\includegraphics[width=1\textwidth]{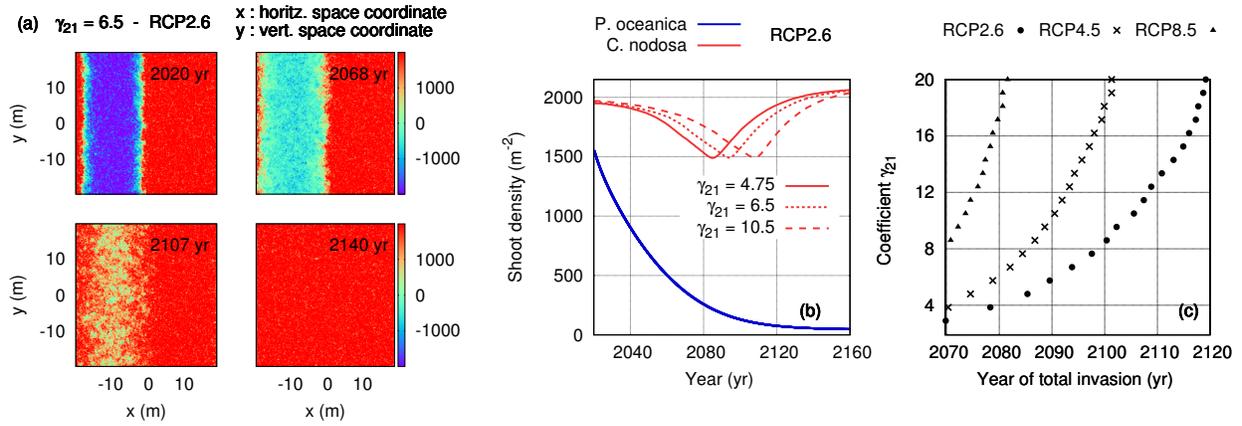}
    \caption{ \label{fig:tempevol} \small Study of the domain separation between {\it Posidonia oceanica} (blue) and {\it Cymodocea nodosa} (red) patches considering the different scenarios of greenhouse emissions. (a) Snapshots are taken between years $2020$ and $2120$ for the case with $\gamma_{21}=6.5$ and with a Representative Concentration Pathway of greenhouse gases corresponding to stringent mitigation measures (RCP2.6) \citep{IPCC2014}. The color bar represents $\rho_{Po}-\rho_{Cn}$, the difference between shoot densities of  {\it P. oceanica} ($\rho_{Po}$) and {\it C. nodosa} ($\rho_{Cn}$) in units of [shoots$\cdot m^{-2}$]. (b) Average shoot density of both species for the selected $\gamma_{21}$ values for the scenario RCP2.6. (c) Coupling coefficient $\gamma_{21}$ vs. the year at which the total occupation by {\it C. nodosa} takes place for the different warming scenarios. For each combination of parameters, the time of total occupation would correspond to the position of the lower peak in the red curve in graphic (a).}
 \end{figure}

Once we have obtained a good estimate for the coupling coefficient $\gamma_{21}$, we incorporate the effect of the variability of the surface seawater temperature on the ecosystem dynamics. 
As mentioned above, {\it C. nodosa} is not negatively affected by seawater temperatures below $34\celsius$ \citep{Savva_2018}, and we can assume a constant mortality rate $\mu_0=0.92 yr^{-1}$ (see Table S1 in the Supplementary Material).  
For  {\it P. oceanica}, its mortality rate is found to increase linearly with time (eq. \ref{mortt}). As a consequence, in the context of global warming, {\it P. oceanica} species must become extinct. The relevant question is to estimate the time to extinction. To do so, we propose a basic mathematical model for the mono-specific case of a single patch of {\it P. oceanica}.
For the sake of simplicity, we neglect the shoot density dependence of the branching rates and species competition. Then, the change in the shoot population can be easily written in terms of a first-order differential equation (a detailed derivation of the mathematical model can be found in the Supplementary Material SM2): 
\be\label{Ndif}
\frac{d N(t)}{dt} =  \left( \nu_0-\mu(t) \right) N(t).
\ee
The solution to this equation predicts that stable seagrass populations of {\it P. oceanica} will decay exponentially as:
\be\label{Nform}
N(t) = N_0 \exp{\left[-{\alpha \over 2}{t^2} + (\nu_0-\mu_0)t\right]}\,,
\ee
where $\nu_0$ is the branching rate, $\mu_0$ the intrinsic mortality rate,  $N_0$ the initial number of shoots, and $\alpha=0.028 \lambda$, with $\lambda$ representing the increase rate of sea-water temperature $[\celsius\cdot yr^{-1}]$. 
The temperatures have been estimated assuming a $SST_{max}=29\celsius$ in the year 2020  in the Balearic Sea with a relative uncertainty of $20\%$  \citep{informeMarBalear2}.
This exponential decay agrees with our simulations when the numerical model includes only single species. These results are summarised in Fig. \ref{fig:exp}. Starting with a patch of {\it P. oceanica} with a uniform distribution of shoots, we have investigated the population dynamics under different global warming scenarios (RCP2.6, RCP4.5, RCP8.5). Interestingly, eq. \eqref{Nform} shows that the decay of the total seagrass population is completely characterised by the clonal growth parameters of single shoots (see Table \ref{table:exp}). The dominant quadratic term is proportional to the factor $\lambda$, which indicates that the regression of {\it P. oceanica} is led by the increase of the seawater temperature. 
\\
\\
The three different emission scenarios that we have simulated predict that stable {\it P. oceanica} meadows will lose $90\%$ of their population by year $2097(\pm4),\,2084(\pm3),$ and $2067(\pm7)$, respectively. At this stage, seagrass meadows would difficulty recover, and their habitats can be considered functionally extinct. Also, in Table \ref{table:tsum}, we show the percentages of {\it P. oceanica} cover expected in the years 2050 and 2100, together with estimated sea-water temperatures. These predictions are derived from the results shown in Fig. \ref{fig:exp}(a) (see also Fig. S1, in the Supplementary Material where errors associated to the uncertainty in the seawater temperature are shown in differents plots for the sake of clarity). In Fig. \ref{fig:exp}(b), we have plotted in a semi-logarithmic scale the change in the number of shoots of  {\it P. oceanica} versus time. Results can be fitted with a quadratic function $f(t) = \left( a{t^2} + b t + c \right)$. The best fit to the coefficients $a,b,c$ for different values of $\lambda$ are summarised in Table \ref{table:exp}, which also shows that they follow the expected relations to recover equation \eqref{Nform}:  $a = - 0.028\, \lambda /2$, $b= (\nu-\mu_0)$ and $c=\log(N_0)$.  Although the mathematical model was deduced ignoring the density carrying capacity ($\epsilon \rightarrow 0$), it still yields a good approximation to predict the time to extinction.
\\
\\
In the following, we investigate how the time to extinction of {\it P. oceanica} meadows is affected by the presence of {\it C. nodosa} and its sensitivity to the value of the coupling parameter $\gamma_{21}$. We will also characterise how {\it C. nodosa} colonises the empty regions left by {\it P. oceanica}.
\\
\\
We start simulating the combined dynamics of {\it P. oceanica} and {\it C. nodosa} assuming the sea-warming predicted by \citep{Darmaraki} in a positively-moderate scenario of emissions (RCP4.5: $\lambda = 2.0\celsius/100yr$). In Fig  \ref{fig:tempevol}, we assume that before year $2020$ {\it P. oceanica} was neither affected by the temperature nor by local anthropogenic disturbances ($\mu_0=0.03 yr^{-1}$). The initial density conditions are taken such that at year $2020$, we have  $40\%$ coverage of {\it P. oceanica} and $60\%$  of {\it C. nodosa} in the study area.
From the year 2020 the mortality rate of {\it P. oceanica}  is replaced by equation \eqref{mortt} with intrinsic mortality $\mu_0=0.07 yr^{-1}$ that accounts for present local disturbances \citep{Jord__2012}. {\it C. nodosa} is  thermally resilient \citep{Savva_2018}, and its mortality rate will be kept constant. 
Different snapshots of the meadow are presented in Fig. \ref{fig:tempevol} for a coupling parameter of $\gamma_{21}=6.5$, fixed to agree with field observations (see Section \ref{sec:comp}).
As a difference to the case of optimal growth conditions (see Fig. \ref{fig:general}), under global warming {\it P. oceanica} is in constant regression and is not able to displace {\it C. nodosa}. Moreover, the average total shoot density of {\it P. oceanica} monotonically decreases, and the shoots of {\it C. nodosa} keep on colonising the space left by the {\it P. oceanica}.
Since the densities in the area of coexistence are lower than in the mono-specific domains, the progressive invasion of the {\it C. nodosa} increases the front width and causes an initial decrease of its average shoot density at the front (Fig. \ref{fig:tempevol}(a), red curve). Remarkably, the location of the minimum in the density of {\it C. nodosa} corresponds to the characteristic time at which the former meadow of {\it P. oceanica} has become colonised by {\it C. nodosa}.
A  sensitivity analysis of the dependence of this characteristic time to the coupling coefficient $\gamma_{21}$ is shown in Fig. \ref{fig:tempevol}(b).
We observe that the time for a total invasion is delayed as we increase the coefficient $\gamma_{21}$. This behaviour is related to Fig. \ref{fig:width}(a,b), where the threshold densities for the coexistence of {\it C. nodosa} and  {\it P. oceanica} are lower for higher values $\gamma_{21}$.
Hereafter, {\it P. oceanica} keeps regressing due to the increase in temperature until its disappearance, leading to the formation of {\it C. nodosa} mono-specific meadows. According to the model outcomes, this situation will occur a couple of years before the {\it P. oceanica} meadows would reach their functional extinction. For a coupling parameter of $\gamma_{21}=6.5$, the model predicts that {\it C. nodosa}  would completely invade {\it P. oceanica} meadows by the year $2094,\,2081,$ and $2065$, for each of the simulated emission scenarios (RCP2.6, RCP4.5, and RCP8.5, respectively). 
%


%

\section{Discussion}
\label{sec:end}

In this work, we have implemented numerical models proposed in  \citep{TS2005,llabres2022mathematical} to study the combined dynamics of {\it P. oceanica} and {\it C. nodosa} species.
These models offer an innovative perspective to understand the most relevant mechanisms behind the growth of seagrasses and, at the same time, allow us to predict their behaviour and spatial distribution under different environmental scenarios, which is the main scope of the present paper.
\\
\\
We have shown that under favorable growth conditions, these species do not mix, but a striped-like region of coexistence appears at the front between patches. 
We have been able to characterise this coexistence region by measuring its width and shoot densities at the front. At the same time, 
field observations performed in Ses Olles de Son Saura (Balearic Islands, Western Mediterranean Sea) have been used to test the model predictions and to fit the coupling parameter $\gamma_{21}\approx 6.5$ that determines how {\it C. nodosa} is affected by {\it P. oceanica}.
Field data also provided evidence of an exponential decay relation between the front width and the length of the {\it P. oceanica} leaves  (Fig. \ref{fig:leafs}(a)). This result, combined with the model outcome that relates the front width with the coupling parameter  $\gamma_{21}$ (Fig. \ref{fig:width}(d)), allows us to derive a linear relationship between the coupling coefficient $\gamma_{21}$ and the leaf length of {\it P. oceanica} (Fig. \ref{fig:leafs}(b)). This is a remarkable result since the coupling parameter between species cannot be directly fitted from field measurements.
\\
\\
We found empirical evidence, at the study site, of unusual short {\it P. oceanica} leaves. The measured average length of $8.15 cm$, which is rather small compared to the $100 cm$ length inside a healthy meadow, makes possible the coexistence region with {\it C. nodosa} at the front. This observation supports the idea that it is the scarce light available to {\it C. nodosa} inside a {\it P. oceanica} meadow one of the main facts that inhibits its growth. Several studies indicate that shorter {\it P. oceanica} leaves could be related to deterioration in the seagrass due to nutrient enrichment, urban effluents, or fish farms \citep{Short:1995aa,DELGADO1999109,Ruiz_2001,Balestri_2004,LEONI20061}. 
Therefore, one could conjecture that the appearance of multi-specific meadows is associated with the degradation of {\it P. oceanica}. This hypothesis, however, should be tested with further experiments.
%
%
\\
\\
The phenomenological origin of our model allows us to understand more profoundly some of the individual mechanisms behind the dynamics of seagrass meadows, in particular, under different environmental scenarios.
We have been able to characterise the exponential decay of the shoot population from purely clonal growth processes when the shoot mortality rate is affected by global warming. Equation \eqref{Nform} describes the regression of {\it P. oceanica} when is affected by an increase in seawater temperature, and Fig. \ref{fig:exp} represents the decaying trends for {\it P. oceanica} predicted by our simulations under different warming scenarios. In a positively-moderate scenario (RCP4.5) of greenhouse emissions, we have estimated that stable meadows will lose $90\%$ of suitable habitat by the year $2084(\pm 3)$. Previous predictions by \cite{Jord__2012} forecasted the functional extinction of {\it P. oceanica} by year $2049$  using a similar warming scenario. The discrepancies between these results have their origin in the form of the exponential behaviour: while \cite{Jord__2012} assumed a decay of the type $N(t) = N_0 \exp{\left[-{\alpha}\,{t^2} + (\nu-\mu_0)t\right]}$, we derived equation \eqref{Nform} purely from the clonal growth rules included in our model. These expressions differ by a factor of two in their exponents leading to faster decay in \cite{Jord__2012}. 
\\
\\
    The relation between global warming and seagrasses has also been studied in \cite{Chefaoui_2018} using ecological niche modelling \citep{elith2009}. Their results predict the geographical redistribution of seagrasses emphasizing the differences between genetic seagrass regions in the Mediterranean Sea \citep{gen1, gen3, gen2}. On the contrary, we use phenomenologically based clonal rules to simulate the seagrasses, fixing the model parameters with field observations from the Western Mediterranean region (Table S1 in the Supplementary Material). For the future sea temperature trends, we have used the available predictions found in the literature, which consist of Mediterranean area averages \citep{Darmaraki}. To forecast more precisely the evolution of seagrasses in the Western Mediterranean, we should have used specific temperature trends in this region, which were not accessible to us at the moment of completing this work. Therefore, we consider the decay rates for {\it P. oceanica} displayed in Figs.\ref{fig:exp}-\ref{fig:tempevol} to be an overestimation for the Western Mediterranean region, which would warm up slower than the Mediterranean average. However, it is remarkable that we obtain similar results to \cite{Chefaoui_2018} when they average over Western, Eastern, and Central Mediterranean regions. Both methods forecast a 70-75$\%$ loss of {\it P. oceanica} by 2050, and its functional extinction by 2100 in the worst-case scenario of emissions (RCP8.5). This match could suggest that the choice of regional growth parameters in our simulations is not as decisive as the use of average temperature. 
 However, some studies find that Mediterranean seagrasses show marked differences in their sensitivity to warming depending on the regionality \citep{marba_resilience, bennet2022b}. 
\\
\\
We have also found that sea warming directly affects the spatial distribution of {\it P. oceanica} and {\it C. nodosa} seabeds. 
While {\it P. oceanica} meadows deteriorate due to warming, {\it C. nodosa} acts as an opportunistic species and conquers the space left by {\it P. oceanica} (Fig. \ref{fig:tempevol}). Remarkably, the time required by {\it C. nodosa} to colonise a {\it P. oceanica} meadow is related to the minimum of the {\it C. nodosa} shoot density in Fig. \ref{fig:tempevol}(a). 
Our results differ from \citep{Chefaoui_2018} that predict a loss of 45$\%$ of {\it C. nodosa} by 2100 in the worst-case scenario of greenhouse emissions (RCP8.5).
Our results are based on the high thermal resilience of {\it C. nodosa}: effective growth happens from  $13.1\celsius$ upwards and becomes optimal between 25-33$\celsius$ \citep{Savva_2018}. On the contrary, in \cite{Chefaoui_2018} correlations between  {\it C. nodosa} and the maximum  $SST$, are implemented in their simulations. The discrepancies between both results will be investigated in prospective work.
\\
\\
Currently, and for simplicity, the mathematical models presented in this work neglect some known mechanisms that seagrasses use to cope with unfavourable conditions and extreme events. Seagrasses, as well as other clonal plants, self-organise in macroscopical spatial patterns to maximise survival. This behavior is modelled by considering non-local interactions among seagrass shoots, as done in \citep{Ruiz-Reynes:2017aa, Ruiz_Reyn_s2_2020}. The analysis of such effects from the numerical perspective could explain the presence of seagrasses in severely warm environments, such as the atoll-like structures of {\it P. oceanica} observed in Western Sicily \citep{Tomasello}. At the physiological level, {\it P. oceanica} responds to warming with massive flowering events \citep{Blok, RUIZ}, and it also possesses a thermal stress memory \citep{Nguyen,PAZZAGLIA}. By overlooking the escape mechanisms in our model, our predictions are underestimating the decaying trends of seagrasses. Also, we will like to emphasize that in our model, we are solely exploring the response to seawater temperature, but other parameters, such as salinity, pH, or nutrients,  might affect the resilience of seagrasses to future climate scenarios. 
\\
\\
Our simulations show that global warming is the leading effect causing the alarming loss of {\it P. oceanica}. The predicted decay rates have still to be validated from time series of shoot densities that were not available to us at the moment of publication and would take years to collect. In any case, the possible extinction of {\it P. oceanica} and other seagrass species will cause dramatic losses in the biodiversity of coastal ecosystems and are expected to have high impacts on their dependent human communities, as a reduction in the total global capacity of carbon sequestration \citep{Waycott:2009aa,DUARTE201332}. 
These concerns are reflected in the Sixth Assessment Report of the Intergovernmental Panel on Climate Change (IPCC) \citep{ipcc2022}, where seagrasses are claimed as one of the main vulnerable ecosystems requiring exceptional preservation. The report emphasizes that limiting warming to 1.5$\celsius$ from pre-industrial levels requires gas emissions to peak before 2025 at the latest and be reduced by 43$\%$ by 2030. Beyond this temperature, irreversible losses will occur in terrestrial and marine ecosystems \citep{IPCC2014}. For {\it P. oceanica}, which has very slow rhizome growing rates (1-6 $cm yr^{-1}$) \citep{10.2307/24857076}, strict preservation measures should be implemented for a complete recovery of the meadows.
Recent studies show that conservationist policy-making effectively reduces the declining trends of seagrass meadows or even reverses them \citep{delossantos, Dunic}. Also, active restoration programs, such as the physical planting of seagrasses or seed distribution, have been very successful for meadow rehabilitation \citep{DuarteRebuilding, Unsworth}, and will also contribute to achieving the total recovery of seagrasses ecosystems.

\section*{ACKNOWLEDGMENTS}

E.LL. and T.S. acknowledge the financial support from projects: PRD2018/18-2 funded by LIET - ITS 2017-006 from the Direcci\'o. Gral. d'Innovaci{\'o} i Recerca (CAIB); PID2021-123723OB-C22 funded by MCIN/AEI/10.13039/501100011033/ ERDF, a way of making Europe and CEX2021-001164-M also funded by MCIN/AEI/10.13039/501100011033, and the Vicen\c c  Mut post-doctoral fellowship funded by ERDF, a way of making Europe and the Conselleria de Fons Europeus, Universitat i Cultura (CAIB). The authors would like to also thank Gabriel Jord{\`a}, N{\'u}ria Marb{\`a}, and Marlene Wesselmann for useful discussions. 



\bibliography{seagrasses}

\newpage

\section*{Figures and tables}

  \begin{table}[h]
\adjustbox{max width=\textwidth}{%
\centering
\begin{tabular}{ccccc}
\hline
  Sea surface warming rate  &  $\%$ cover in 2050 & $SST_{max}$ in 2050 &  $\%$ cover in 2100 & $SST_{max}$ in 2100\\  \hline
 (RCP2.6) \quad  1.0 $\celsius/100yr$ &  $48\pm10\%$ & $29.3\celsius$ & $8\pm2\%$ & $29.8\celsius$\\
 (RCP4.5) \quad 2.0 $\celsius/100yr$ &  $42\pm9\% $  & $29.6\celsius$ & $3\pm1\%$ & $30.6\celsius$\\
 (RCP8.5)  \quad  4.5 $\celsius/100yr$ &  $30\pm6\% $ & $30.4\celsius$ & $0.4\pm0.1\%$ &$32.6\celsius$\\
\end{tabular}}
\caption{\label{table:tsum}\small Estimated percentage of coverage of {\it Posidonia oceanica} meadows in year $2050$ and $2100$ under different warming scenarios compared to year 2020. The sea surface warming rates are Mediterranean area averages corresponding to the predictions in \citep{Darmaraki}. The temperatures have been estimated assuming $SST_{max} = 29\celsius$ in the year 2020, as measured in the Balearic Sea region  \citep{informeMarBalear2}, and have a relative uncertainty of 20$\%$. }
 \end{table}

\qquad
 \qquad
 
 \begin{table}[h]
\adjustbox{max width=\textwidth}{%
\centering
\begin{tabular}{cccccccc}
\hline
   $\lambda$   & $a$ & $b$ &  $c$ & - $\alpha /2$  & $\nu-\mu_0$ &$\log(N_0)$\\ \hline
   $\celsius/100yr$ & yr$^{-2}$& yr$^{-1}$ & adim. & yr$^{-2}$& yr$^{-1}$& adim. \\
 1.0  & $(-1.59\pm 0.02)\cdot 10^{-4}$   & $-0.017\pm 0.003 $  & $13.793 \pm 0.001$ &  $-1.54\cdot 10^{-4}$  & $-0.02$ & 13.85  \\
 2.0 & $(-2.98\pm 0.03)\cdot 10^{-4}$   & $-0.0185\pm 0.0004 $  & $13.812 \pm 0.001$ &  $-3.0\cdot 10^{-4}$  & id. & id.  \\
 4.5 & $(-6.31\pm 0.03)\cdot 10^{-3}$   & $-0.0202\pm 0.0001 $  & $13.817 \pm 0.001$ &  $-6.3\cdot 10^{-3}$  & id. & id.  \\
\end{tabular}}
\caption{\label{table:exp}\small Parameters derived from the  exponential fit  $f(t)=a{t^2} + b t + c$ done in Fig. \ref{table:exp}(b). The comparison
between the coefficients $a,b,c$ and the model parameters (see last three columns) show an excellent agreement between the simulation results and the analytical solution of the model (see Supplementary Material S2).}
 \end{table}

\qquad
 \qquad

\qquad
 \qquad
 
 \qquad
 \qquad
 
 
 \qquad
 \qquad
 
 \qquad
 \qquad
 
   \qquad
 \qquad
 
\end{document}